# Tuning Non-Gilbert-type damping in FeGa films on MgO(001) via oblique deposition

Yang Li[1,2], Yan Li[1,2], Qian Liu[3], Zhe Yuan[3], Qing-Feng Zhan[4], Wei He[1], Hao-Liang Liu[1], Ke Xia[3], Wei Yu[1], Xiang-Qun Zhang[1], Zhao-Hua Cheng[1,2,5 a)]

[1]State Key Laboratory of Magnetism and Beijing National Laboratory for Condensed Matter Physics, Institute of Physics, Chinese Academy of Sciences, Beijing 100190, China

[2]School of Physical Sciences, University of Chinese Academy of Sciences, Beijing 100049, China

[3]The Center for Advanced Quantum Studies and Department of Physics, Beijing Normal University, 100875 China

[4]State Key Laboratory of Precision Spectroscopy, School of Physics and Materials Science, East China Normal University, Shanghai 200241, China

[5]Songshan Lake Materials Laboratory, Dongguan, Guangdong 523808, China

a) Corresponding author, e-mail: zhcheng@iphy.ac.cn

**Abstract**

The ability to tailor the damping factor is essential for spintronic and spin-torque applications. Here, we report an approach to manipulate the damping factor of FeGa/MgO(001) films by oblique deposition. Owing to the defects at the surface or interface in thin films, two-magnon scattering (TMS) acts as a non-Gilbert damping mechanism in magnetization relaxation. In this work, the contribution of TMS was characterized by in-plane angular dependent ferromagnetic resonance (FMR). It is demonstrated that the intrinsic Gilbert damping is isotropic and invariant, while the extrinsic mechanism related to TMS is anisotropic and can be tuned by oblique deposition. Furthermore, the two and fourfold TMS related to the uniaxial magnetic anisotropy (UMA) and magnetocrystalline anisotropy were discussed. Our results open an avenue to manipulate magnetization relaxation in spintronic devices.

# 1. Introduction

In the past decades, controlling magnetization dynamics in magnetic nanostructures has been extensively studied due to its great importance for spintronic and spin-torque applications [1,2]. The magnetic relaxation is described within the framework of the Landau-Lifshitz Gilbert (LLG) phenomenology using the Gilbert damping factor α [3]. The intrinsic Gilbert damping depends primarily on the spin-orbit coupling (SOC) [4,5]. It has been demonstrated that alloying or doping with non-magnetic transition metals provides an opportunity to tune the intrinsic damping [6,7]. Unfortunately, in this way the soft magnetic properties will reduce. In addition to the intrinsic damping, the two-magnon scattering (TMS) process serves as an important extrinsic mechanism in magnetization relaxation in ultrathin films due to the defects at surface or interface [8,9]. This process describes the scattering between the uniform magnons and degenerate final-state spin wave modes [10]. The existence of TMS has been demonstrated in many systems of ferrites [11-13]. Since the anisotropic scattering centers, the angular dependence of the extrinsic TMS process exhibits a strong in-plane anisotropy [14], which allows us to adjust the overall magnetic relaxation, including both the intensity of relaxation rate and the anisotropic behavior.

Here, we report an approach to engineer the damping factor of $Fe_{81}Ga_{19}$ (FeGa) films by oblique deposition. The FeGa alloy exhibits large magnetostriction and narrow microwave resonance linewidth [15], which could assure it as a promising material for spintronic devices. For the geometry of off-normal deposition, it has been demonstrated to provoke shadow effects and create a periodic stripe defect matrix. This can introduce a strong uniaxial magnetization anisotropy (UMA) perpendicular to the projection of the atom flux [16-19]. Even though some reports have shown oblique deposition provokes a twofold TMS channel [20-22], the oblique angle dependence of the intrinsic

Gilbert damping and the TMS still remain in doubt. For our case, on the basis of the first-principles calculation and the in-plane angular-dependent FMR measurements, we found that the intrinsic Gilbert damping is isotropic and invariant with varying oblique deposition angles, while the extrinsic mechanism related to the two-magnon-scattering (TMS) is anisotropic and can be tuned by oblique deposition. In addition, importantly we firstly observe a phenomenon that the cubic magnetocrystalline anisotropy determines the area including degenerate magnon modes, as well as the intensity of fourfold TMS. In general, the strong connection between the extrinsic TMS and the magnetic anisotropy, as well their direct impact on the damping constants, are systemically investigated, which offer us a useful approach to tailor the damping factor.

## 2. Experimental details

FeGa thin films with a thickness of 20 nm were grown on MgO(001) substrates in a magnetron sputtering system with a base pressure below $3 \times 10^{-7}$ Torr. Prior to deposition, the substrates were annealed at 700 °C for 1 h in a vacuum chamber to remove surface contaminations and then held at 250 °C during deposition. The incident FeGa beam was at different oblique angles of $\psi$=0°, 15°, 30°, and 45°, with respect to the surface normal, and named S1, S2, S3, and S4 in this paper, respectively. The projection of FeGa beam on the plane of the substrates was set perpendicular to the MgO[110] direction, which induces a UMA perpendicular to the projection of FeGa beam, i.e., parallel to the MgO[110] direction, due to the well-known self-shadowing effect. Finally, all the samples were covered with a 5 nm Ta capping layer to avoid surface oxidation [see figure 1(a)]. The epitaxial relation of FeGa(001)[110]||MgO(001)[100] was characterized by using the X-ray in-plane Φ-scans, as described elsewhere [23]. Magnetic hysteresis loops were measured at various in-plane magnetic field orientations $\varphi_H$ with respect to the FeGa [100] axis using magneto-optical Kerr effect (MOKE) technique at room temperature. The dynamic magnetic properties were investigated by broadband FMR measurements based on a broadband vector network analyzer (VNA) with a transmission geometry coplanar waveguide (VNA-FMR) [24]. This setup allows both frequency and field-sweeps measurements with external field applied parallel to the sample plane. During measurements, the samples were placed face down on the coplanar waveguide and the transmission coefficient $S_{21}$ was recorded.

## 3. Results and discussion

Figure 1(b) displays the Kerr hysteresis loops of sample S1 and S4 recorded along with the main crystallographic directions of FeGa[100], [110], and [010]. The sample S1 exhibits rectangular hysteresis curves with small coercivities for the magnetic field along [100] and [010] easy axes. In contrast, the S4 displays a hysteresis curve with two steps for the magnetic field along the [010] axis, which indicates a UMA along the FeGa[100] axis superimposed on the fourfold magnetocrystalline anisotropy. As a result, with increasing the oblique angle, the angular dependence of normalized remnant magnetization ($M_r/M_s$) gradually reveals a fourfold symmetry combined with a uniaxial symmetry, as shown in the inset of figure 1(b).

Subsequently, the magnetic anisotropic properties can be further precisely characterized by the in-plane angular-dependent FMR measurements. Figure 1(c) and 1(d) show typical FMR spectra for the real and imaginary parts of coefficient $S_{21}$ for the sample S2. Recorded FMR spectra contain a symmetric and an antisymmetric Lorentzian peak, from which the resonant field $H_r$ with linewidth $\Delta H$ can be obtained [24,25].

Figure 2(a) shows the in-plane angular dependence of $H_r$ measured at 13.0 GHz and can be fitted by the following expression [26,27]:

$$f = \frac{\gamma\mu_0}{2\pi}\sqrt{H_a H_b} \qquad (1)$$

Here, $H_a = H_4(3 + cos4\varphi_M)/4 + H_u cos^2\varphi_M + H_r cos(\varphi_M - \varphi_H) + M_{eff}$ and $H_b = H_4 cos4\varphi_M + H_u cos2\varphi_M + H_r cos(\varphi_M - \varphi_H)$, $H_4$ and $H_u$ represent the fourfold anisotropy field and the UMA field caused by the self-shadowing effect, respectively. $\varphi_H(\varphi_M)$ is the azimuthal angles of the applied field (the tipped magnetization) with respect to the [100] direction, as depicted in figure 1(a). $\mu_0 M_{eff} = \mu_0 M_s - \frac{2K_{out}}{M_s}$, $M_s$ is the saturation magnetization and $K_{out}$ is the out-of-plane uniaxial anisotropy constant. $f$ is the resonance frequency, $\gamma$ is the gyromagnetic ratio and here used as the accepted value for Fe films, $\gamma$=185 rad GHz/T [28].

The angular dependent $H_r$ reveals only a fourfold symmetry for the none-obliquely deposited sample, which indicates the cubic lattice texture of FeGa on MgO. With increasing the oblique angle, a uniaxial symmetry is found to be superimposed on the fourfold symmetry, clearly confirming a UMA is produced by the oblique growth, which agrees with the MOKE' results. The fitted parameter $\mu_0 M_{eff} = 1.90 \pm 0.05$T is found to be independent on the oblique deposition and close to $\mu_0 M_s = 1.89 \pm 0.02$ T estimated using VSM, which is almost same as the value of the literature [29]. This indicates negligible out-of-plane magnetic anisotropy in the thick FeGa films. As shown in figure 2(b), it is observed that the UMA ($K_u=H_u M_s/2$) exhibits a general increasing trend with oblique angle, which coincides with the fact the shadowing effect is stronger at larger angles of incidence [16-19]. Interestingly the oblique deposition also affects the cubic anisotropy $K_4$ ($K_4=H_4 M_s/2$). Different from the $K_4$ increases slightly with deposition angle in Co/Cu system [16], here the value of $K_4$ is the lowest at an oblique angle of 15°. It is well known that film stress significantly influences the crystallization tendency [30,31]. FeGa alloy is highly stressed sensitive due to its larger magnetostriction. Thus, the change in $K_4$ of FeGa films may be attributed to the anisotropy dispersion created due to the stress variations during grain growth. It should be mentioned that the best way to determine magnetic parameters is to measure the out-of-plane FMR. But the effective saturation magnetization $\mu_0 M_{\text{eff}} =$ 1.90T of FeGa alloy leads to the perpendicular applied field beyond our instrument limit. Meanwhile, the results obtained above are also in accord with those extracted by fitting field dependence of the resonance frequency with H//FeGa[100] shown in figure 2(c).

The effective Gilbert damping $\alpha_{\text{eff}}$ is extracted by linearly fitting the dependence of linewidth on frequency: $\mu_0 \Delta H = \mu_0 \Delta H_0 + \frac{2\pi f \alpha_{\text{eff}}}{\gamma}$, where $\Delta H_0$ is the inhomogeneous broadening. For the sake of clarity, figure 3(a) only shows the frequency dependence of linewidth for the samples S1 and S2 along [110] and [100] axes. It is evident that, for the sample S1, both linear slopes of two directions are almost same. While with regard to the sample S2, the slope of the $\Delta H$-$f$ curve along the easy axis is approximately a factor of 2 greater than that of the hard axis. The obtained values of $\alpha_{\text{eff}}$ are shown in figure 3(b). Firstly, the results clearly indicate that the effective damping exhibits anisotropy, with higher value along the easy axis. Secondly, for the easy axis, the oblique angle dependence on the damping parameter indicates an extraordinary trend and has a peak at deposition angle 15°. However, the damping shows an increasing trend with the oblique angle for the field along the hard axis. In the following part, we will explore the effect of oblique deposition on the mechanism of the anisotropic damping and the magnetic relaxation process.

So far, convincing experimental evidence is still lacking to prove the existence of anisotropic damping in bulk magnets. Chen *et al*. have shown the emergence of anisotropic Gilbert damping in ultrathin Fe (1.3nm)/GaAs and its anisotropy disappears rapidly when the Fe thickness increases [32]. We perform the first-principles calculation of the Gilbert damping of FeGa alloy considering the effect induced by the lattice distortion. We artificially make a tetragonal lattice with varying the lattice constant of the c-axis. The electronic structure of Fe-Ga alloy is calculated self-consistently using the coherent potential approximation implemented with the tight-binding linear muffin-tin orbitals. Then the atomic potentials of Fe and Ga are randomly distributed in a 5×5 lateral supercell, which is connected to two semi-infinite Pd leads. A thermal lattice disorder is included via displacing atoms randomly from the perfect lattice sites following a Gaussian type of distribution [33]. The root-mean-square displacement at room temperature is determined by the Debye model with the Debye temperature 470 K. The length of the supercell is variable and the calculated total damping is scaled linearly with this length. Thus, a linear least-squares fitting can be performed to extract the bulk damping of the Fe-Ga alloy [34]. The calculated Gilbert damping is plotted in figure 3(c) as a function of the lattice distortion $(c-a)/a$. The Gilbert damping is nearly independent of the lattice distortion and there is no evidence of anisotropy in the intrinsic bulk damping of FeGa alloy.

So the extrinsic contributions are responsible for the anisotropic behavior of damping, which can be separated from the in-plane angular dependent linewidth. The recorded FMR linewidth have the following different contributions [11]:

$$\mu_0 \Delta H = \mu_0 \Delta H^{\mathrm{inh}} + \frac{2\pi \alpha_G f}{\gamma \Phi} + \left| \frac{\partial H_\mathrm{r}}{\partial \varphi_\mathrm{H}} \Delta \varphi_\mathrm{H} \right| + \sum_{<x_i>} \Gamma_{<x_i>} f(\varphi_\mathrm{H} - \varphi_{<x_i>}) \arcsin \sqrt{(\sqrt{\omega^2 + (\frac{\omega_0}{2})^2} - \frac{\omega_0}{2})/(\sqrt{\omega^2 + (\frac{\omega_0}{2})^2} + \frac{\omega_0}{2})} + \Gamma^{\mathrm{max}}_{\mathrm{twofold}} cos^4(\varphi_\mathrm{M} - \varphi_{\mathrm{twofold}}) \quad (2)$$

$\Delta H^{\mathrm{inh}}$ is both frequency and angle independent term due to the sample inhomogeneity. The second term is the intrinsic Gilbert damping ($\alpha_G$) contribution. $\Phi$

is a correction factor owing to the field dragging effect caused by magnetic anisotropy [12], $\Phi = cos\,(\varphi_\mathrm{M}-\varphi_\mathrm{H})$. The $\varphi_\mathrm{M}$ as a function of $\varphi_\mathrm{H}$ for the sample S2 at fixed 13 GHz is calculated and shown in figure 4(a). Note that the dragging effect vanishes ($\varphi_\mathrm{M} = \varphi_\mathrm{H}$) when the field is along the hard or easy axes. The third term describes the mosaicity contribution originating from the angular dispersion of the crystallographic cubic axes and yields a broader linewidth [35]. The fourth term is the TMS contribution. The $\Gamma_{<x_i>}$ signifies the intensity of the TMS along the principal in-plane crystallographic direction $<x_i>$. The $f(\varphi_\mathrm{H}-\varphi_{<x_i>})$ term indicates the TMS contribution depending on the in-plane direction of the field relative to $<x_i>$ and commonly expressed as $cos^2[2(\varphi_\mathrm{M}-\varphi_{<x_i>})]$ [14]. In addition, $\omega$ is the angular resonant frequency and $\omega_0 = \gamma\mu_0 M_\mathrm{eff}$. In our case, besides the fourfold TMS caused by expected lattice geometric defects, the other twofold TMS channel is induced by the dipolar fields emerging from periodic stripelike defects [20,21]. This term is parameterized by its strength $\Gamma_\mathrm{twofold}^\mathrm{max}$ and the axis of maximal scattering rate $\varphi_\mathrm{twofold}$.

As an example, the angle-dependent linewidth measured at 13.0 GHz for the sample S2 is shown in figure 4(b). It clearly exhibits a strong in-plane anisotropy, and the linewidth along the [100] direction is significantly larger than that along the [110] direction. Taking only isotropic Gilbert damping into account, the dragging effect vanishes with field applied along the hard and easy axes. Meanwhile, the mosaicity term gives an angular variation of the linewidth proportional to $|\partial H_r/\partial \varphi_H|$, which is also zero along with the principal <100> and <110> directions. This gives direct evidence that the relaxation is not exclusively governed only considering the intrinsic Gilbert mechanism and mosaicity term. Because the probability of defect formation along with <100> directions is higher than that along the <110> directions [12], the TMS contribution is stronger along the easy axes, which is in accordance with the fact that the linewidths along the [100] and [110] directions are non-equivalent. Moreover, the linewidth of [010] direction is slightly larger than that along the [100] direction, suggesting that another twofold TMS channel is induced by oblique deposition. As indicated by the red solid line in figure 4(b), the linewidth can be well fitted. Different parts making sense to the linewidth can therefore be separated and summarized in Table I. As we know, the TMS predicts the curved non-linear frequency dependence of linewidth, which not appear in a small frequency range for our case (as shown in figure 3(a)). The linewidth as function of frequency was also well fitted including the TMS-damping using the parameters in Table I (not shown here).

The larger strength of TMS along the easy axis can clearly explain the anisotropic behavior of damping, with higher value along the easy axis shown in figure 3(b). The obtained Gilbert damping factor of ~ $7\times10^{-3}$ is isotropic and invariant with different oblique angles. The value of damping is slightly larger than the bulk value of $5.5\times10^{-3}$ [29], which may be attributed to spin pumping of the Ta capping layer.

The obtained maxima of twofold TMS exhibits an increasing trend with the oblique angle [shown in figure 4(c)]. According to previous works on the shadowing effect [16-19], the larger deposition angle makes the shadowing effect stronger, and the dipolar fields within stripelike defects increase just like the UMA. This can clearly explain that the intensity of twofold TMS follows exactly the same trend with the deposition angle as the UMA. The axis of the maximal intensity of twofold TMS is parallel to the projection of the FeGa atom flux from the fitting data. As shown in Table I, amazingly the modified growth conditions also influence the fourfold TMS, especially the strength of TMS along the <100> axis. Figure 4(c) also presents the changes of the fourfold TMS intensity as the deposition angle and shows a peak at 15°, which follows a similar trend as that of $\alpha_{\text{eff}}$ along [100] axis as shown in figure 3(b). This indeed confirms TMS-damping plays an important role in FeGa thin films.

For the dispersion relation $\omega(k_\parallel)$ in thin magnetic films, the propagation angle $\varphi_{\vec{k}_\parallel}$ defined as the angle between $\vec{k}_\parallel$ and the projection of the saturation magnetization $M_s$ into the sample plane is less than the critical value: $\varphi_{\max} = sin^{-1}\sqrt{\mu_0 H_r/(\mu_0 H_r + \mu_0 M_{\text{eff}})}$ [9,36,37]. This implies no degenerate modes are available for the angle $\varphi_{\vec{k}_\parallel}$ larger than $\varphi_{\max}$. Based on this theory, we propose a hypothesis that the crystallographic anisotropy determines the area including degenerate magnon modes, as well as the intensity of the fourfold TMS. The resonance field along <100> axis changes due to the various crystallographic anisotropy, which has a great effect on the $\varphi_{\max}$. The values of $\varphi_{\max}$ of samples are shown in figure 4(d). The data follow the same trend with the oblique angle as $\Gamma_{<100>}$. During the grain growth, the cubic anisotropy is influenced possibly since the anisotropy dispersion due to the stress. For the lower anisotropy of sample S2, a relatively larger amount of stress and defects present in the sample and lead to a larger fourfold TMS.

## 4. Conclusions

In conclusion, the effects of oblique deposition on the dynamic properties of FeGa thin films have been investigated systematically. The pronounced TMS as non-Gilbert damping results in an anisotropic magnetic relaxation. As the oblique angle increases, the magnitude of the twofold TMS increases due to the larger shadowing effect. Furthermore, the cubic anisotropy dominates the area including degenerate magnon modes, as well as the intensity of fourfold TMS. The reported results confirm that the modified anisotropy can influence the extrinsic relaxation process and open an avenue to tailor magnetic relaxation in spintronic devices.

**Acknowledgments**

This work is supported by the National Key Research Program of China (Grant Nos. 2015CB921403, 2016YFA0300701, and 2017YFB0702702), the National Natural Sciences Foundation of China (Grant Nos. 91622126, 51427801, and 51671212) and the Key Research Program of Frontier Sciences, CAS (Grant Nos. QYZDJ-SSW-JSC023, KJZD-SW-M01 and ZDYZ2012-2). The work at Beijing Normal University is partly supported by the National Natural Sciences Foundation of China (Grant Nos. 61774017, 61704018, and 11734004), the Recruitment Program of Global Youth Experts and the Fundamental Research Funds for the Central Universities (Grant No. 2018EYT03).

**Figure Captions**

**Figure 1** (color online) (a) Schematic illustration of the film deposition geometry and coordinate system (b) In-plane hysteresis loops of samples S1 and S4 with the field along [100], [110], and [010]. The inset shows the polar plot of the normalized remanence ($M_r/M_s$) as a function of the in-plane angle. FMR spectrum for the sample S2 with $H$ along [100] and [110] axes showing the real (c) and imaginary (d) parts of the $S_{21}$.

**Figure 2** (color online) (a) $H_r$ vs. $\varphi_H$ for FeGa films. (b) The anisotropy constants $K_4$ and $K_u$ vs. deposition angle. (c) $f$ vs. $H_r$ plots measured at $H//[100]$, Symbols are experimental data and the solid lines are the fitted results.

**Figure 3** (color online) (a) $\Delta H$ as a function of $f$ for samples S1 and S2 with field along easy and hard axis. (b) The dependence of the damping parameter on the oblique angle with field along [100] and [110] directions. (c) The calculated damping of FeGa alloy as a function of lattice distortion.

**Figure 4** (color online) (a) $\varphi_M$ and (b) $\Delta H$ as a function of $\varphi_H$ for the sample S2 measured at 13.0 GHz. (c) Oblique angle dependences of $\Gamma_{<100>}$ and $\Gamma_{twofold}^{max}$. (d) The largest angle including degenerate magnon modes as a function of the oblique angle with the applied field along <100> direction.

**Table Caption**

Table I. The magnetic relaxation parameters of the FeGa films prepared via oblique deposition (with experimental errors in parentheses).

**Figure 1**

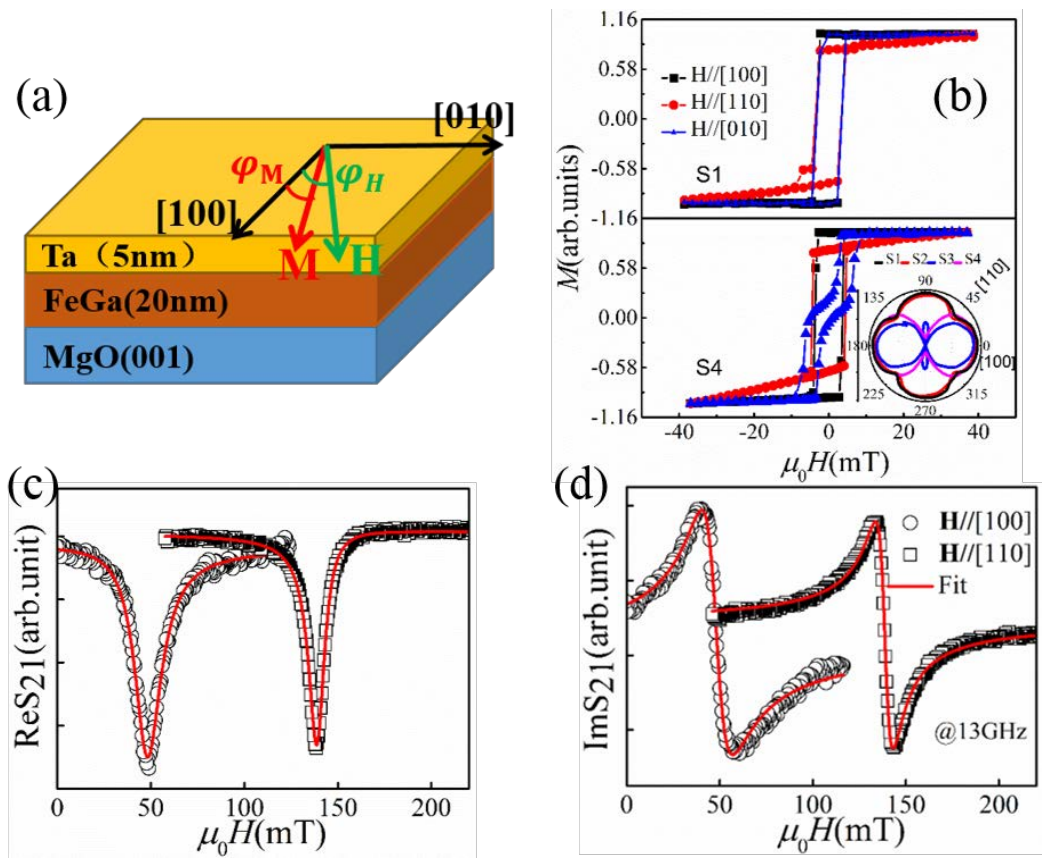

**Figure 2**

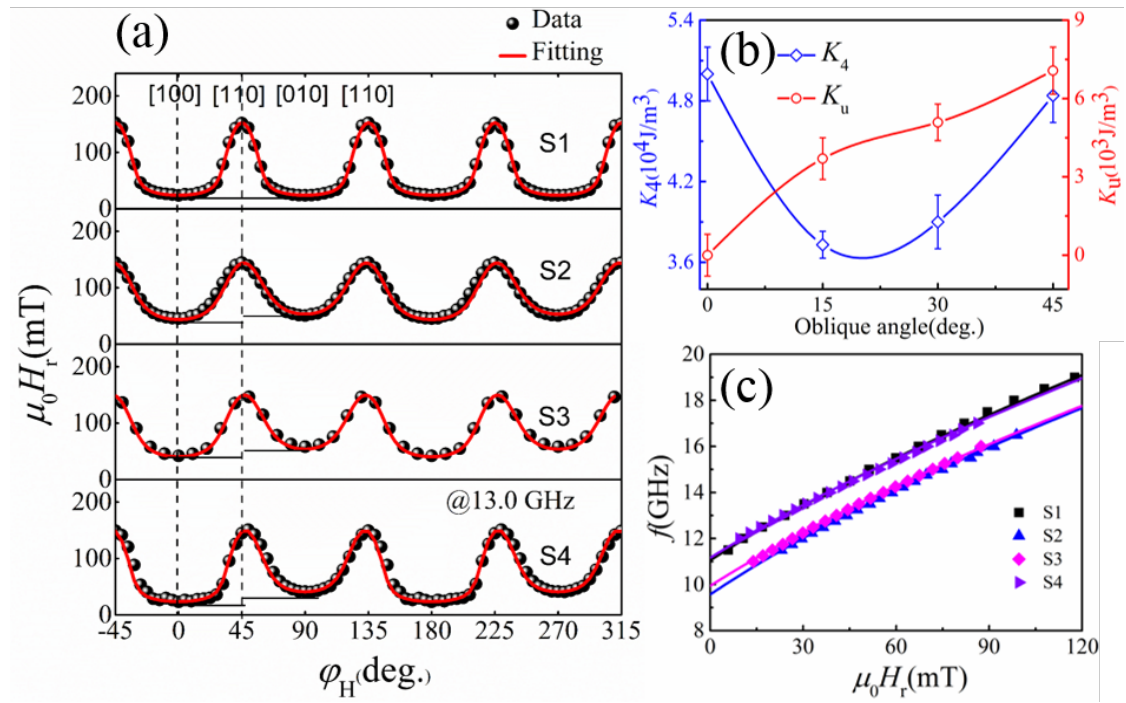

**Figure 3**

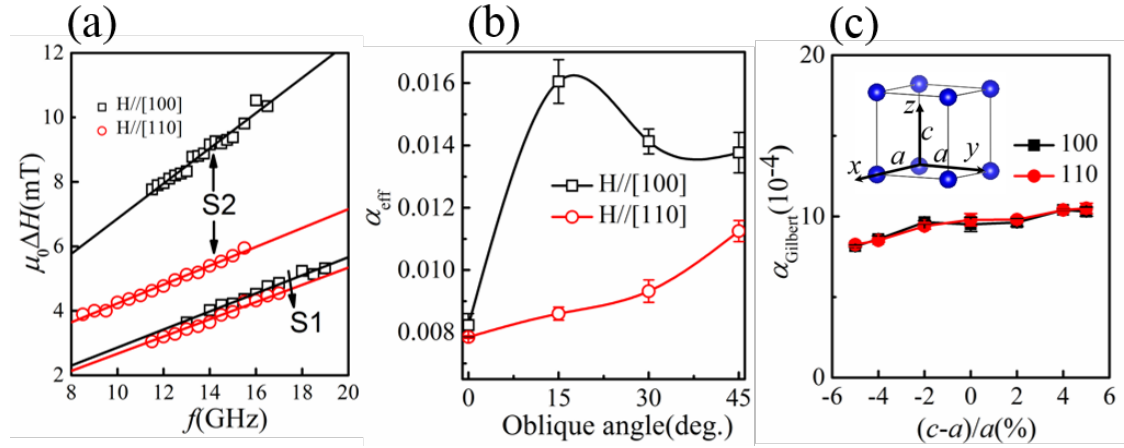

**Figure 4**

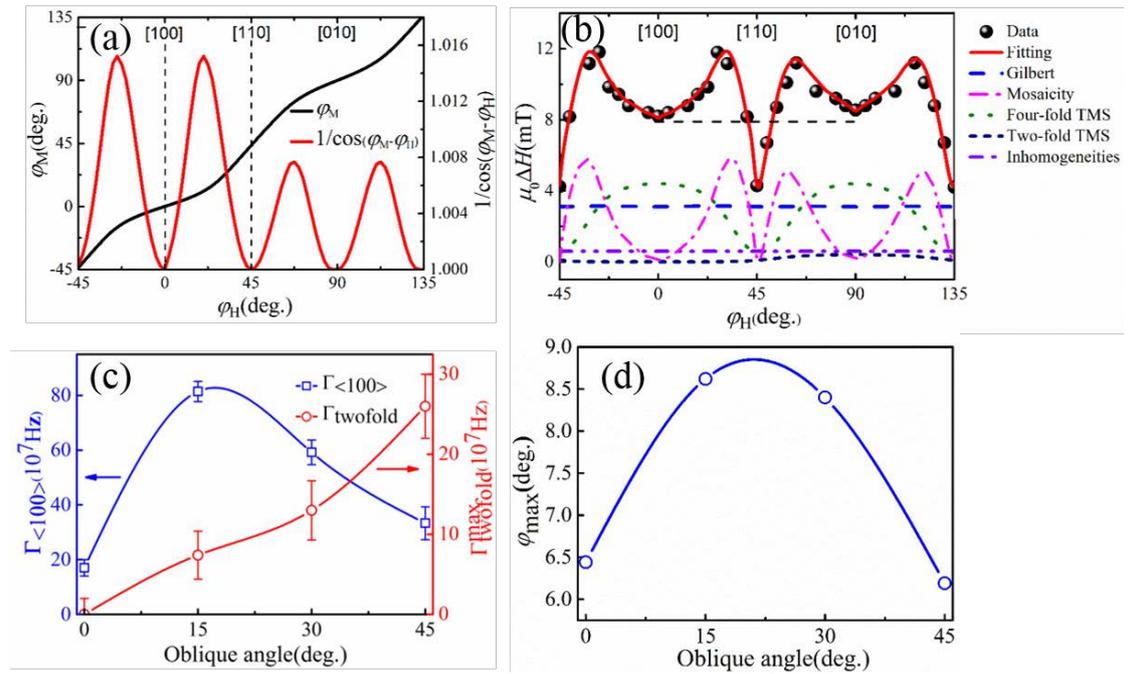

**Table I**

| Sample | $\mu_0 \Delta H^{inh}$ (mT) | $\alpha_G$ | $\Delta \varphi_H$ (deg.) | $\Gamma_{<100>}$ ($10^7$Hz) | $\Gamma_{<110>}$ ($10^7$Hz) | $\Gamma_{twofold}^{max}$ ($10^7$Hz) | $\varphi_{twofold}$ (deg.) |
|---|---|---|---|---|---|---|---|
| S1 | 0 | 0.007 | 0.62 | 17(3) | 5.8(1.8) | 0(2) | 90 |
| S2 | 0.7 | 0.007 | 1.2 | 81.4(3.7) | 9.3(1.9) | 7.4(3) | 90 |
| S3 | 0 | 0.007 | 1.0 | 59.2(4.5) | 11.1(2) | 13(3.7) | 90 |
| S4 | 0 | 0.007 | 1.1 | 33.3(6) | 14.8(3.7) | 26(4) | 90 |